\documentclass{PoS}
\PoS{PoS(LAT2005)347}

\usepackage{graphicx}

\newcommand{\ba}{\begin{eqnarray}}
\newcommand{\ea}{\end{eqnarray}}
\newcommand{\be} {\begin{equation}}
\newcommand{\ee} {\end{equation}}
\newcommand{\GeV}{\mbox{\rm GeV}}
\newcommand{\order}{{\cal O}}

\title{Dynamical determination of $B_K$ from improved staggered quarks }

\ShortTitle{Dynamical determination of $B_K$ from improved staggered quarks }

\author{\speaker{Elvira G\'amiz}, Sara Collins, Christine T.H.~Davies\\
        Department of Physics \&
               Astronomy, University of Glasgow, Glasgow, G12 8QQ, UK\\
        E-mail: \email{e.gamiz@physics.gla.ac.uk}, 
\email{c.davies@physics.gla.ac.uk}}

\author{Junko Shigemitsu\\
Physics Department, The Ohio State
        University, Columbus, OH 43210, USA\\
        E-mail: \email{shige@mps.ohio-state.edu}}

\author{Matthew Wingate\\
Institute for Nuclear Theory, University of
        Washington, Seattle, WA 98115, USA\\
        E-mail: \email{wingate@phys.washington.edu}}

\abstract{The scaling corrections that affected previous staggered calculations
of $B_K$ have been proved to be reduced by using improved
actions (HYP, Asqtad) in the quenched approximation. This improved
behaviour allows us to perform a reliable dynamical calculation
of $B_K$ including quark vacuum polarization effects using the
MILC (2+1) flavour dynamical configurations.
We report here on the results from such dynamical calculation.
We also discuss the renormalization effects with the Asqtad action.}

\FullConference{XXIIIrd International Symposium on Lattice Field Theory\\
		 25-30 July 2005\\
		 Trinity College, Dublin, Ireland}

\begin{document}

\section{Introduction}

In the last few years lattice calculations have started achieving the level of 
precision and the control of uncertainties necessary to extract 
phenomenologically relevant results \cite{Davies:2003ik}. Simulations with 
dynamical quarks are required for this task, since the uncontrolled errors 
associated with the quenched approximation are usually the main source of 
uncertainty in these calculations. An example of this fact can be seen in the 
study of indirect CP violation in the neutral kaon system.

The CP violating effects in $K^0-\bar K^0$ mixing are parametrized by 
$\varepsilon_K$. Experimentally, this quantity is known with a 
few per-cent level precision. On 
the other hand, theoretically it is given by the hadronic matrix 
element between $K^0$ and $\bar K^0$ of the $\Delta S=2$ effective hamiltonian
\be\label{effHam}
H_{eff}^{\Delta S=2}=C_{\Delta S=2}(\mu)\int d^4x \,Q_{\Delta S=2}(x)
\ee
with
\be
Q_{\Delta S=2}(x) = \left[\bar s_{\alpha}\gamma_{\mu}
d_{\alpha}\right]_{V-A}(x)
\left[\bar s_{\beta}\gamma^{\mu}d_{\beta}\right]_{V-A}(x).
\ee

The Wilson coefficient $C_{\Delta S=2}(\mu)$ is a perturbative 
quantity known to 
NLO in $\alpha_s$ in both NDR and HV schemes. It depends on several CKM matrix 
elements about which we would like to obtain information. The matrix element 
$\langle \bar K^0\vert Q_{\Delta S=2}\vert K^0\rangle$, that encodes the 
non-perturbative physics of the problem, is usually normalized by 
its VIA value, defining $B_K$ as the ratio
\be
B_K(\mu) \equiv \frac{\langle\bar K^0|
Q_{\Delta S=2}(\mu)|K^0\rangle}
{\frac{8}{3}\langle \bar K^0|\bar s\gamma_{\mu}\gamma_5d|0\rangle
\langle 0|\bar s\gamma_{\mu}\gamma_5d|K^0\rangle}.
\ee

The main source of uncertainty when one tries to constrain the value of the 
combination of CKM matrix elements involved in the theoretical calculation of 
$\varepsilon_K$ using its experimental value is the error associated to the 
determination of $B_K$ \cite{Charles:2004jd}.  
Improvement in the calculation of $B_K$ 
is thus crucial in order to get information about the unitarity triangle.

The value of $B_K$ that the phenomenologists are using at present in 
their studies of the unitarity triangle \cite{Charles:2004jd}, that is 
considered as the benchmark of the lattice calculations of this parameter, 
was obtained by the JLQCD collaboration \cite{JLQCD97} using unimproved 
staggered quarks in the quenched approximation. The value 
given in \cite{JLQCD97} is $B_K^{NDR}(2~\GeV)=0.628(42)$. 

The main source of uncertainty in this calculation is 
the unknown error from quenching, 
that could be as large as a 15\% according to the ChPT estimate performed in 
\cite{Sharpe97}. In order to have a prediction at a few per-cent level it is thus 
necessary to perform a dynamical calculation of $B_K$ 
that eliminates the quenched 
uncertainties. Another drawback of the calculation in \cite{JLQCD97} 
is the fact that it is affected by large scaling uncertainties. 
We will see that the scaling 
behaviour is going to be much better using improved\footnote{
In the improved actions the thin links are substituted by fattened 
links and the quark-gluon interactions that violate the taste 
symmetry are reduced.} staggered actions instead of the 
standard unimproved staggered action used by the JLQCD collaboration. A third 
way of improving the JLQCD calculation would be incorporating $SU(3)$ 
breaking effects by using kaons made up of non-degenerate quarks, 
instead of degenerate quarks with $m_s/2$.

The goal of this work is to perform a dynamical calculation of $B_K$ that 
eliminates the quenched uncertainty, using improved staggered 
fermions that have been proved to reduce the large $\order(a^2)$ 
discretization errors generated by the taste-changing interactions. 

Preliminary results for that study were presented in \cite{Gamiz:2004qx}. The 
matching coefficients needed in the calculation of the renormalized 
$B_K$ with the action used in our dynamical simulations were 
not available at that moment, so 
an approximate renormalization was performed in order to get those preliminary 
results. The correct renormalization coefficients were calculated later 
\cite{BGM05} and have been used to 
obtain the results reported in these proceedings.

\section{Improved versus unimproved staggered actions: scaling violations}

\label{imversusunim}

The first thing we analyse is the impact of using improved staggered actions 
in comparison with the unimproved staggered action used in \cite{JLQCD97}, that 
suffers from large scaling violations. We do this analysis for two different 
improved actions: the HYP \cite{HYP} and the Asqtad\footnote
{In our quenched analysis we are using a variation of the Asqtad 
action with no improvement in the gauge action.} \cite{Asqtad}.

\begin{table}
\begin{center}
\begin{tabular}{c c c c c}\hline\hline
$\beta$ & Volume & $n_{confs}$ & $a^{-1}(\GeV)$ & $m_s/2$ \\
\hline
5.7 & $12^3\times 24$ & 150 & 0.837(6) & 0.086/0.064 \\
5.93 & $16^3\times 32$ & 50 & 1.59(3) & 0.039/0.030 \\
\hline
\hline
\end{tabular}
\end{center}
\caption{Parameters in the quenched simulations. The values of $m_s/2$ 
are for the HYP and Asqtad staggered actions respectively.
\label{tablequenched}}
\end{table}

The values of the parameters used in the simulations are shown in Table 
\ref{tablequenched}. We choose these parameters to be the same as those used 
by the JLQCD collaboration in order to make a clear comparison 
with their results.  
In particular, we match kaon masses at a given $\beta$ to those of the JLQCD 
collaboration. The matching of the lattice operators to the continuum ones in 
the $\overline{MS}$ scheme have been performed perturbatively 
using the one-loop coefficients calculated in \cite{BGM05}.

\begin{figure}[t]
\begin{center}
\includegraphics [angle=-90,width=85mm] {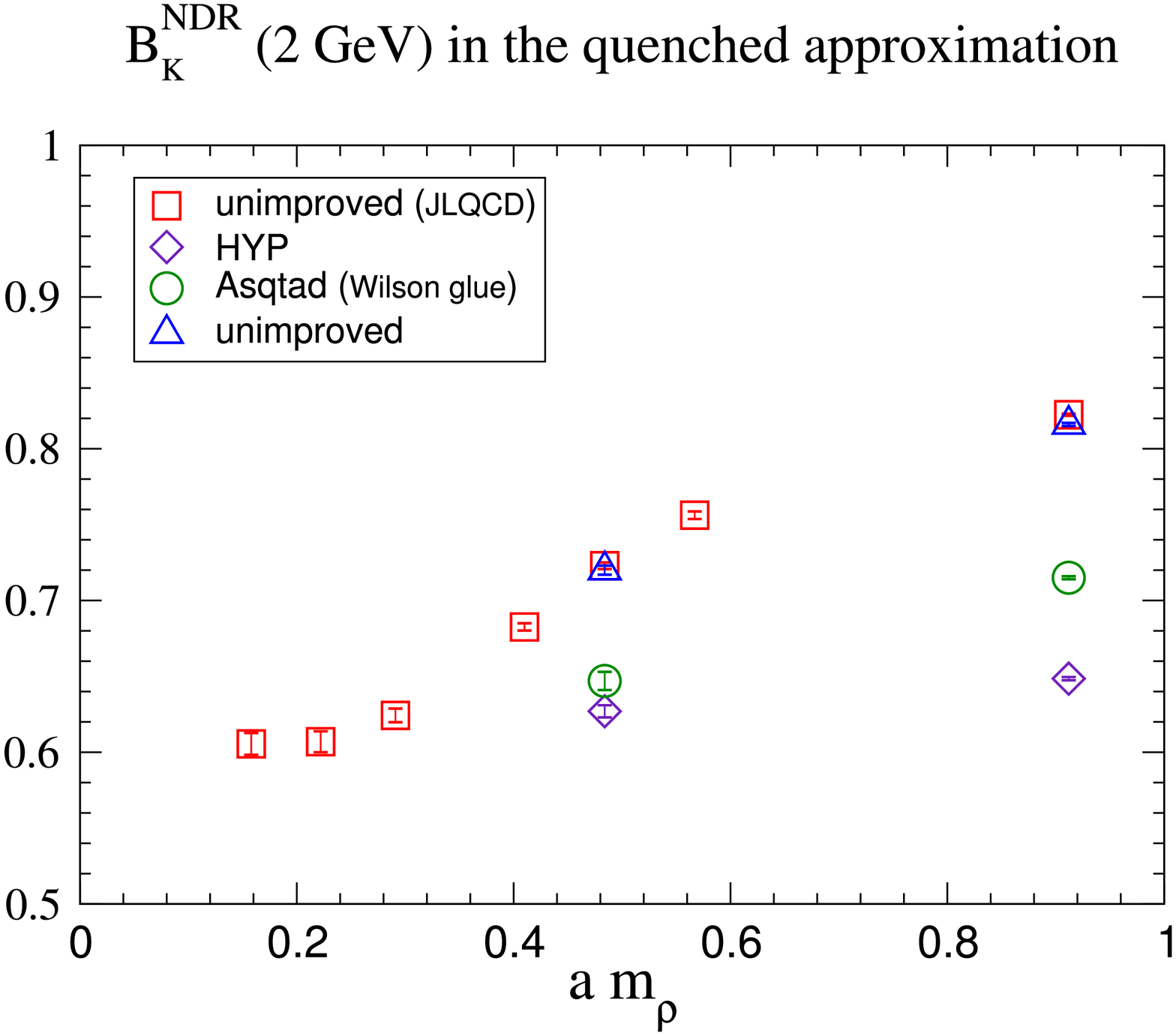}
\end{center}
\caption{Scaling of $B_K^{NDR}(2\GeV)$ with $a$ for improved staggered actions 
compared to the JLQCD unimproved staggered results. \label{quenchedresults}}
\end{figure}

The results we obtain for $B_K^{NDR}(2\GeV)$ as a function 
of the lattice spacing are shown in Figure \ref{quenchedresults}. 
In that Figure a clear improvement in the scaling can be seen when 
using improved actions, in particular in the HYP case. We expect such 
improved scaling to survive unquenching and this therefore allows 
us to perform 
reliable dynamical calculations with only a few values of the lattice 
spacing and even obtain valuable information from a single point simulation.

\section{Dynamical value of $B_K$}

We now incorporate dynamical effects in the calculation of $B_K$ using one of 
the improved staggered actions analyzed in the quenched approximation, Asqtad,
 since the final goal is to eliminate the irreducible systematic 
error associated to quenching that dominates the final uncertainty 
in previous determinations of $B_K$. 

We performed a dynamical calculation of $B_K$ with the Asqtad action, 
using the 
configurations from the MILC collaboration with $n_f=2+1$ dynamical flavours 
\cite{MILC01}. The results reported in these proceedings correspond 
to the analysis 
at one lattice spacing with $a=0.125~{\rm fm}$ and two 
different values of the sea light quark masses. The parameters used in the 
dynamical simulations are collected in Table \ref{tabledyn}.
\begin{table}
\begin{center}
\begin{tabular}{c c c c c}\hline\hline
$\beta$ & $n_{confs}$ & $m_{sea}$ & $m_s/2$& $\alpha_V(1/a)$\\
\hline
6.76 &560 & 0.01/0.05 & 0.02 & 0.4723 \\
6.79 &414 & 0.02/0.05 & 0.02 & 0.4699 \\
\hline
\hline
\end{tabular}
\end{center}
\caption{Parameters in the dynamical simulations.\label{tabledyn}}
\end{table}

The conversion of the values of the bare lattice operators to a value for 
$B_K^{NDR}(2\GeV)$ has been done perturbatively using the $\order(\alpha_s)$ 
lattice to continuum matching coefficients in \cite{BGM05}. 
In the matching process 
we take $\alpha_s$ in the V scheme at the scale $1/a$ where $a$ is the lattice 
spacing -see values in Table \ref{tabledyn}. 
The value for $\alpha_V$ with $N_f=3$ has been taken from the recent 
4-loops lattice determination in \cite{Mason:2005zx}.

\begin{figure}[t]
\begin{center}
\includegraphics [angle=-90,width=85mm] {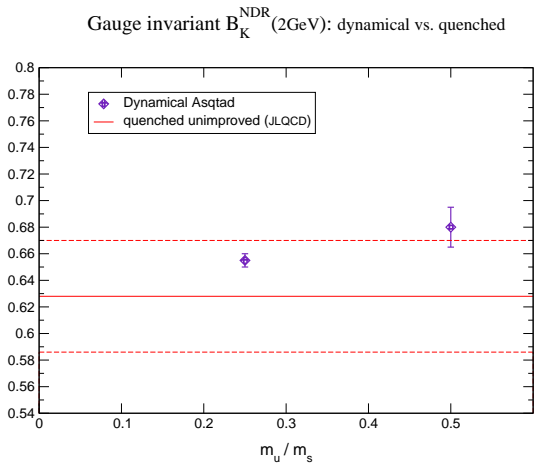}
\end{center}
\caption{Dynamical value of $B_K^{NDR}(2\GeV)$ as a function of 
the ratio between 
the light sea quark mass and the (real) strange quark mass. 
The lines represent 
the quenched results from \cite{JLQCD97}.\label{dynresults}}
\end{figure}

The results we obtain for $B_K^{NDR}(2\GeV)$ including only 
statistical errors as 
a function of the light sea quark masses are shownin Figure \ref{dynresults}.
A decrease of the value of $B_K$ with the reduction of the 
dynamical quark mass can 
be appreciated in that Figure. After linearly extrapolating these 
results to the 
physical $s$ and $u(d)$ masses following the staggered chiral 
perturbation study in 
\cite{SW05}, the value of $B_K^{NDR}(2\GeV)$ we obtain is
\be\label{bkvalue}
B_K^{NDR}(2\GeV)=0.618(18)(19)(30)(130)\, ,
\ee
where the first error is statistical, the second is from chiral fits, 
the third one is from discretizations errors and the final one is from the 
perturbative conversion 
to the $\overline{MS}$ scheme. The value in (\ref{bkvalue}) is equivalent to 
$\hat B_K=0.83\pm0.18$, with $\hat B_K$ defined as the product 
of $B_K^{NDR}(2\GeV)$ 
and the Wilson coefficient in the effective hamiltonian (\ref{effHam}) in the 
$\overline{MS}-NDR$ scheme and at 2$\GeV$. This quantity is scheme and scale 
independent at $\order(\alpha_s)$. Note that this value of $\hat B_K$ is very 
similar to the previous quenched staggered result in \cite{JLQCD97} 
($\hat B_K=0.86(6)(14)$), so any final conclusion about the enhancement or  
decreasing due to the inclusion of dynamical effect in its calculation would 
need a drastic reduction of the error quoted in (\ref{bkvalue}). That error 
is dominated by the uncertainty associated to the possible 
$\order(\alpha_s^2)$ 
corrections in the lattice to continuum matching process, that we estimate 
to be of the order of $\left(\alpha_V(1/a)\right)^2$. 
At the lattice spacing we 
are working, the unquenched values of $\alpha_V$ are given in Table 
\ref{tabledyn} and translate into a $\sim 20\%$ error in the result for $B_K$.

\section{Summary and discussion}

Most of the previous lattice calculations of $B_K$, in particular that used in 
the unitarity triangle analysis, were performed in the quenched approximation. 
This induces a large, 
essentially unknown and irreducible systematic error into 
the result. Precise simulations with dynamical fermions are necessary in order 
to be able to make full use of the experimental data on $\varepsilon_K$ to 
constrain the CKM matrix. These dynamical 
simulations are feasible with present 
computers using staggered fermions at light dynamical quark masses. However, 
the unimproved staggered action suffers from large taste-changing interactions 
that generate important scaling corrections, as those found by the JLQCD 
results. We have shown in Section \ref{imversusunim} that the scaling 
behaviour is much better when using improved staggered actions, that have 
been designed to reduce the non-physical taste-changing 
interactions as well as other cut-off effects. 

As a first step in the dynamical study of $B_K$ on the MILC 
configurations we have 
calculated this quantity in two ensembles at $a=0.125~{\rm fm}$ and 
with two different 
light sea quark masses. In doing that, we have used the recent results for the 
matching coefficients corresponding to the Asqtad action \cite{BGM05}, the one 
used in the numerical simulations. The value of the renormalized $B_K$ 
in the $\overline{MS}$ scheme we obtain is $B_K^{NDR}(2\GeV)=0.618\pm0.136$ or,
 equivalently, $\hat B_K=0.83\pm0.18$. No sizeable deviation  
from quenched results can be inferred from this preliminary value. 
The uncertainty in this determination is dominated by the perturbative 
error associated to the one-loop matching process, 
so, in order to have the precision of a few per-cent needed by phenomenology, 
a two-loop matching or a numerical matching method is required.

The next step will be to redo our calculation on a finer lattice 
\cite{CDGSW05}. This new result for a different lattice spacing will 
allow us to perform an appropriate continuum extrapolation\footnote
{In view of the improved scaling behaviour we obtained in the quenched 
approximation within the Asqtad action described in 
Section \ref{imversusunim}, results for two lattice spacings will 
be enough to perform a reliable continuum limit.}, 
reducing the final discretization errors. 
The perturbative error will also be reduced since $\alpha_V(1/a)$ 
is smaller for finer lattices. And we will be able to check whether 
the good scaling behaviour observed in the quenched approximation 
using improved staggered actions (Asqtad) is in fact present in the 
dynamical results.

We also want to investigate other issues in the study of $B_K$, 
like the differences 
between using invariant and non-invariant gauge operators, the application of 
the staggered chiral perturbation theory results in \cite{SW05}, the impact of 
$SU(3)$ breaking effects or the chiral limit value of this quantity 
that could be compared to recent continuum calculations \cite{Bkchiral}.

\section{Acknowledgments} 

E.G. is indebted to the European Union for a Marie 
Curie Intra-European Fellowship. The work is also supported by PPARC and DoE.


\begin{thebibliography}{99}

\bibitem{Davies:2003ik}
C.~T.~H.~Davies {\it et al.}  [HPQCD Collaboration],
Phys.\ Rev.\ Lett.\  {\bf 92} (2004) 022001 
[hep-lat/0304004].

\bibitem{Charles:2004jd}
  J.~Charles {\it et al.}  [CKMfitter Group],
  Eur.\ Phys.\ J.\ C {\bf 41} (2005) 1
  [hep-ph/0406184];
  M.~Bona {\it et al.}  [UTfit Collaboration],
J. High Energy Phys. {\bf 0507} (2005) 028 
[hep-ph/0501199].

\bibitem{JLQCD97} S.~Aoki {\it et al.}  [JLQCD Collaboration],
Phys.\ Rev.\ Lett.\  {\bf 80} (1998) 5271
[hep-lat/9710073].

\bibitem{Sharpe97} S.~R.~Sharpe,
Nucl.\ Phys.\ Proc.\ Suppl.\  {\bf 53} (1997) 181
[hep-lat/9609029].

\bibitem{Gamiz:2004qx}
  E.~G\'amiz, S.~Collins, C.~T.~H.~Davies, J.~Shigemitsu and M.~Wingate  [HPQCD
                  Collaboration],
   Nucl.\ Phys.\ Proc.\ Suppl.\  {\bf 140} (2005) 353 
  [hep-lat/0409049].


\bibitem{BGM05}
T.~Becher, E.~G\'amiz and K.~Melnikov, 
Phys.\ Rev.\ D, accepted for publication [hep-lat/0507033].

\bibitem{HYP}
A.~Hasenfratz and F.~Knechtli,
Phys.\ Rev.\ D {\bf 64} (2001) 034504
[hep-lat/0103029].

\bibitem{Asqtad}
S.~Naik,
Nucl.\ Phys.\ B {\bf 316} (1989) 238; 
G.~P.~Lepage,
Phys.\ Rev.\ D {\bf 59} (1999) 074502
[hep-lat/9809157]; 
K.~Orginos, D.~Toussaint and R.~L.~Sugar  [MILC Collaboration],
Phys.\ Rev.\ D {\bf 60} (1999) 054503
[hep-lat/9903032]. 

\bibitem{MILC01} C.~W.~Bernard {\it et al.},
Phys.\ Rev.\ D {\bf 64} (2001) 054506
[hep-lat/0104002].

\bibitem{Mason:2005zx}
  Q.~Mason {\it et al.}  [HPQCD Collaboration],
  Phys.\ Rev.\ Lett.\  {\bf 95} (2005) 052002
  [hep-lat/0503005].

\bibitem{SW05}
R.~S.~Van de Water and S.~R.~Sharpe,
 hep-lat/0507012, see also these proceedings.

\bibitem{CDGSW05}
  S.~Collins, C.~T.~H.~Davies, E.~G\'amiz, J.~Shigemitsu and M.~Wingate  [HPQCD
                  Collaboration], in preparation.

\bibitem{Bkchiral}
J.~Prades, J.~Bijnens and E.~G\'amiz,
 hep-ph/0501177; 
 O. Cat\`a and S. Peris,
J. High Energy Phys.  {\bf 03} (2003) 060  
[hep-ph/0303162]; 
S. Peris and E. de Rafael,
Phys.\ Lett.\ B {\bf 490} (2000) 213 
[Erratum, hep-ph/0006146]; 
J. Bijnens and J. Prades,
J. High Energy Phys. {\bf 01} (2000) 002 
[hep-ph/9911392].


\end{thebibliography}
\end{document}